\input harvmac
\let\includefigures=\iftrue
\let\useblackboard=\iftrue
\newfam\black 

\includefigures
\message{If you do not have epsf.tex (to include figures),}
\message{change the option at the top of the tex file.}
\input epsf
\def\figin{\epsfcheck\figin}\def\figins{\epsfcheck\figins}
\def\epsfcheck{\ifx\epsfbox\UnDeFiNeD
\message{(NO epsf.tex, FIGURES WILL BE IGNORED)}
\gdef\figin##1{\vskip2in}\gdef\figins##1{\hskip.5in}
\else\message{(FIGURES WILL BE INCLUDED)}%
\gdef\figin##1{##1}\gdef\figinbs##1{##1}\fi}
\def\DefWarn#1{}
\def\figinsert{\goodbreak\midinsert}
\def\ifig#1#2#3{\DefWarn#1\xdef#1{fig.~\the\figno}
\writedef{#1\leftbracket fig.\noexpand~\the\figno}%
\figinsert\figin{\centerline{#3}}\medskip\centerline{\vbox{
\baselineskip12pt\advance\hsize by -1truein
\noindent\footnotefont{\bf Fig.~\the\figno:} #2}}
\endinsert\global\advance\figno by1}
\else
\def\ifig#1#2#3{\xdef#1{fig.~\the\figno}
\writedef{#1\leftbracket fig.\noexpand~\the\figno}%
\global\advance\figno by1} \fi

\def\id{{1 \kern-.28em {\rm l}}}

\def\K3{{\bf K3}}
\def\journal#1&#2(#3){\unskip, \sl #1\ \bf #2 \rm(19#3) }
\def\andjournal#1&#2(#3){\sl #1~\bf #2 \rm (19#3) }

\def\bar{\overline}
\def\hat{\widehat}
\def\ie{{\it i.e.}}
\def\eg{{\it e.g.}}

\def\tilde{\widetilde}

\def\frac#1#2{{#1\over#2}}

\def\inbar{\,\vrule height1.5ex width.4pt depth0pt}
\def\IC{\relax\hbox{$\inbar\kern-.3em{\rm C}$}}
\def\IR{\relax{\rm I\kern-.18em R}}
\def\IZ{\relax{\rm I\kern-.18em Z}}

%
%

%
\catcode`\@=11
\def\slash#1{\mathord{\mathpalette\c@ncel{#1}}}
\overfullrule=0pt

\def\NN{{\cal N}}

\def\WW{{\cal W}}

\def\underrel#1\over#2{\mathrel{\mathop{\kern\z@#1}\limits_{#2}}}

\catcode`\@=12


%


\def\ie{{\it i.e.}}
\def\eg{{\it e.g.}}

\lref\AlmuhairiWS{
  A.~Almuhairi and J.~Polchinski,
  ``Magnetic $AdS x R^2$: Supersymmetry and stability,''
[arXiv:1108.1213 [hep-th]].
}

\lref\SeibergPQ{
  N.~Seiberg,
  ``Electric - magnetic duality in supersymmetric nonAbelian gauge theories,''
Nucl.\ Phys.\ B {\bf 435}, 129 (1995).
[hep-th/9411149].
}

\lref\WittenYC{
  E.~Witten,
  ``Phases of N=2 theories in two-dimensions,''
Nucl.\ Phys.\ B {\bf 403}, 159 (1993).
[hep-th/9301042].
}

\lref\KutasovNP{
  D.~Kutasov and A.~Schwimmer,
  ``On duality in supersymmetric Yang-Mills theory,''
Phys.\ Lett.\ B {\bf 354}, 315 (1995).
[hep-th/9505004].
}

\lref\KutasovSS{
  D.~Kutasov, A.~Schwimmer and N.~Seiberg,
  ``Chiral rings, singularity theory and electric - magnetic duality,''
Nucl.\ Phys.\ B {\bf 459}, 455 (1996).
[hep-th/9510222].
}

\lref\KutasovVE{
  D.~Kutasov,
  ``A Comment on duality in N=1 supersymmetric nonAbelian gauge theories,''
Phys.\ Lett.\ B {\bf 351}, 230 (1995).
[hep-th/9503086].
}

\lref\KutasovIY{
  D.~Kutasov, A.~Parnachev and D.~A.~Sahakyan,
  ``Central charges and U(1)(R) symmetries in N=1 superYang-Mills,''
JHEP {\bf 0311}, 013 (2003).
[hep-th/0308071].
}

\lref\BrodieVX{
  J.~H.~Brodie,
  ``Duality in supersymmetric SU(N(c)) gauge theory with two adjoint chiral superfields,''
Nucl.\ Phys.\ B {\bf 478}, 123 (1996).
[hep-th/9605232].
}

\lref\IntriligatorMI{
  K.~A.~Intriligator and B.~Wecht,
  ``RG fixed points and flows in SQCD with adjoints,''
Nucl.\ Phys.\ B {\bf 677}, 223 (2004).
[hep-th/0309201].
}

\lref\MartinecZU{
  E.~J.~Martinec,
  ``Algebraic Geometry and Effective Lagrangians,''
Phys.\ Lett.\ B {\bf 217}, 431 (1989).
}

\lref\VafaUU{
  C.~Vafa and N.~P.~Warner,
  ``Catastrophes and the Classification of Conformal Theories,''
Phys.\ Lett.\ B {\bf 218}, 51 (1989).
}

\lref\GiveonSR{
  A.~Giveon and D.~Kutasov,
  ``Brane dynamics and gauge theory,''
Rev.\ Mod.\ Phys.\  {\bf 71}, 983 (1999).
[hep-th/9802067].
}

\lref\GaddeLXA{
  A.~Gadde, S.~Gukov and P.~Putrov,
  ``(0,2) Trialities,''
[arXiv:1310.0818 [hep-th]].
}

\lref\BeniniCZ{
  F.~Benini and N.~Bobev,
  ``Exact two-dimensional superconformal R-symmetry and c-extremization,''
Phys.\ Rev.\ Lett.\  {\bf 110}, no. 6, 061601 (2013).
[arXiv:1211.4030 [hep-th]].
}
\lref\BeniniCDA{
  F.~Benini and N.~Bobev,
  ``Two-dimensional SCFTs from wrapped branes and c-extremization,''
JHEP {\bf 1306}, 005 (2013).
[arXiv:1302.4451 [hep-th]].
}

\lref\KutasovUX{
  D.~Kutasov,
  ``New results on the 'a theorem' in four-dimensional supersymmetric field theory,''
[hep-th/0312098].
}

\lref\KutasovFFL{
  D.~Kutasov and J.~Lin,
  ``(0,2) Dynamics From Four Dimensions,''
[arXiv:1310.6032 [hep-th]].
}

\lref\KutasovYQA{
  D.~Kutasov and J.~Lin,
  ``Exceptional N=1 Duality,''
[arXiv:1401.4168 [hep-th]].
}

\Title{}
{\vbox{\centerline{(0,2) ADE Models From Four Dimensions}
\bigskip
}}
\bigskip

\centerline{\it David Kutasov and Jennifer Lin}
\bigskip
\centerline{EFI and Department of Physics, University of
Chicago} \centerline{5640 S. Ellis Av., Chicago, IL 60637, USA }
\smallskip

\vglue .3cm

\bigskip

\let\includefigures=\iftrue
\bigskip
\noindent 
Four dimensional $\NN=1$ supersymmetric gauge theories with unitary gauge groups and matter in the adjoint and fundamental representations give rise to a series of non-trivial fixed points with an ADE classification. Many of these models exhibit generalizations of Seiberg duality. Upon compactification on a two-torus with suitable background fields for global $U(1)$ symmetries, they flow at long distances to two dimensional theories with $(0,2)$ supersymmetry. We study these theories, focusing on the two dimensional analogs of the four dimensional dualities.

\bigskip

\Date{}

\newsec{Introduction}

In a recent paper \KutasovFFL, we studied a general construction that associates a two dimensional quantum field theory with $(0,2)$ supersymmetry to a four dimensional $\NN=1$ supersymmetric field theory equipped with a global  $U(1)$ symmetry. This construction involves a compactification of the four dimensional theory on a two-torus in the presence of a constant background magnetic field and auxiliary $D$-field for the global $U(1)$, that together preserve half of the supersymmetry \AlmuhairiWS. 

In \KutasovFFL\ we applied this procedure to supersymmetric QCD with gauge group $U(N_c)$ coupled to $N_f$ fundamental chiral superfields $Q^i$, $\tilde Q_i$.  We considered the case of even $N_f$ and took the $U(1)$ global symmetry that figures in the construction to be the subgroup of the  $SU(N_f) \times SU(N_f)$ global symmetry group, $U(1)_e$, that assigns charge $+1$ to half of the $Q$'s and $\tilde Q$'s and $-1$ to the other half.\foot{For other choices of the global $U(1)$ symmetry, supersymmetry is typically broken in the quantum theory.} We saw that in this case the theory flows in the infrared to a non-trivial $(0,2)$ SCFT, and described some of its properties. 

In particular, we investigated the effect of compactification on the Seiberg duality of the four dimensional theory. It is a priori not obvious whether the four dimensional duality descends to a two dimensional one. In \KutasovFFL\ we assumed that this is the case, and found a picture consistent with that assumption. At the same time, it would be nice to provide further evidence for it. 

The aim of this note is to take a step in this direction, by generalizing the discussion of \KutasovFFL\ to a larger class of four dimensional theories, which contain in addition to the fields listed above two chiral superfields $X$, $Y$ transforming in the adjoint representation of the $U(N_c)$ gauge group. As was pointed out in \IntriligatorMI, this class of models gives rise to a large set of non-trivial four dimensional $\NN=1$ SCFT's with an ADE classification. The different theories are labeled by the superpotential of the adjoint fields, $W(X,Y)$, which takes the form 
\eqn\rgfp{
\matrix{
 \hat O & W_{\hat O} = 0 \cr
 \hat A & \;\;\;\;\;\;\;W_{\hat A} = \Tr Y^2 \cr
 \hat D & \;\;\;\;\;\;\;\;\;\;W_{\hat D} = \Tr XY^2\cr
 \hat E & \;\;\;\;\;\;\;W_{\hat E} = \Tr Y^3 \cr
 A_k & \;\;\;\;\;\;\;\;\;\;\;\;\;\;\;\;\;\;\;\;\;\;\;\;W_{A_k} = \Tr(X^{k+1}+Y^2) \cr
 D_{k+2} & \;\;\;\;\;\;\;\;\;\;\;\;\;\;\;\;\;\;\;\;\;\;\;\;\;\;\;\;\;\;\;W_{D_{k+2}} = \Tr (X^{k+1}+XY^2) \cr
 E_6 & \;\;\;\;\;\;\;\;\;\;\;\;\;\;\;\;\;\;\;\;W_{E_6} = \Tr (Y^3 + X^4) \cr
 E_7 & \;\;\;\;\;\;\;\;\;\;\;\;\;\;\;\;\;\;\;\;\;\;\;W_{E_7} = \Tr (Y^3 + YX^3) \cr
 E_8 & \;\;\;\;\;\;\;\;\;\;\;\;\;\;\;\;\;\;\;\;\;\;W_{E_8} = \Tr(Y^3 + X^5)\,.
}}
For each theory there are upper and lower bounds on the number of flavors of fundamental superfields for which the corresponding fixed points exist. For the $A_k$ theories these were found in \refs{\KutasovVE\KutasovNP\KutasovSS-\KutasovIY}, for $D_k$ in \BrodieVX, and for $E_7$ in \KutasovYQA. The general situation was discussed in \IntriligatorMI. 

The $A$ and $D$ series theories exhibit generalizations of Seiberg duality \refs{\KutasovVE,\KutasovNP,\KutasovSS,\BrodieVX}. We recently proposed an analog of these dualities for the $E_7$ case \KutasovYQA.
One of our goals is to investigate the consequences of these four dimensional dualities for the low energy two dimensional $(0,2)$ theories obtained via the construction of \KutasovFFL. In the next two sections we discuss in turn the $A$ and $D$ series models. In section 4 we comment on the generalization of the construction to the other theories in \rgfp. 

We find that the construction of \KutasovFFL\ generalizes naturally to this larger class of theories. We view this as  further evidence  for the picture proposed in that paper.

\newsec{$A$ series}

The starting point of our discussion is a four dimensional $\NN=1$ supersymmetric gauge theory  with gauge group $U(N_c)$, coupled to $N_f$ superfields  in the (anti) fundamental representation of  the gauge group, $Q^i, \tilde Q_i$, $i=1,2,\cdots, N_f$, and an adjoint chiral superfield $X$. The $A_k$ theory in \rgfp\  is obtained by turning on the superpotential\foot{The second adjoint field $Y$ is massive in this case, and thus can be omitted without changing the infrared behavior.}  
\eqn\epot{
\WW = g\Tr X^{k+1}.
}
For $k=1$, the superpotential \epot\ is a mass term for $X$, which thus decouples in the IR, giving ordinary supersymmetric QCD. For $k=2$, the coupling $g$ \epot\ is marginal. It becomes relevant in the presence of the gauge interaction when the theory is asymptotically free (\ie\ for $N_f<2N_c$), and induces a flow to a new fixed point  \KutasovVE. For $k>2$, \epot\ is superficially irrelevant, but it becomes relevant in the presence of the gauge interaction when the number of flavors is sufficiently small \refs{\KutasovNP,\KutasovIY,\KutasovUX}.  This gives an upper bound on the number of flavors for which the $A_k$ fixed point exists (or, more precisely, is distinct from the $\hat A$ one). 

The lower bound on $N_f$ is due to the fact that the $A_k$ theory only has a vacuum if \refs{\KutasovVE,\KutasovNP}
\eqn\stabv{
N_f \geq \frac{N_c}{k}.
}
One way to see this is to deform the superpotential \epot\ to an arbitrary polynomial of degree $k+1$, $\WW(x)$, with the same large $x$ behavior. Since this deformation does not change the leading behavior of the potential at infinity, it is not expected to change the answer to the question of whether the theory has supersymmetric vacua. 

For generic $\WW$, the potential for $X$ has $k$ distinct minima, the solutions of the F-term equation $W'(x)=0$. Vacua are obtained by using gauge symmetry and the D-term constraints to diagonalize $X$, and distributing the $N_c$ eigenvalues among the different minima. Classically there are many such vacua, in each of which the adjoint field is massive and the low energy theory splits into $\NN=1$ SQCD theories corresponding to different minima of the potential. The number of colors in each minimum is given by the number of eigenvalues of $X$ placed in that minimum, while the number of flavors in all of them is $N_f$.  

Quantum mechanically, $\NN=1$ SQCD has a vacuum only when the number of flavors is larger or equal than the number of colors.  Imposing this constraint on the theories associated with the individual minima implies that in the deformed theory, and therefore in the undeformed one as well, a vacuum exists if and only if \stabv\ is satisfied (see \refs{\KutasovNP,\KutasovSS} for more detailed discussions).

\bigskip
\vbox{
$$\vbox{\offinterlineskip
\hrule height 1.1pt
\halign{&\vrule width 1.1pt#
&\strut\quad#\hfil\quad&
\vrule width 1.1pt#
&\strut\quad#\hfil\quad&
\vrule width 1.1pt#
&\strut\quad#\hfil\quad&
\vrule width 1.1pt#
&\strut\quad#\hfil\quad&
\vrule width 1.1pt#
&\strut\quad#\hfil\quad&
\vrule width 1.1pt#
&\strut\quad#\hfil\quad&
\vrule width 1.1pt#\cr
height3pt
&\omit&
&\omit&
&\omit&
&\omit&
&\omit&
&\omit&
\cr
&\hfil field&
&\hfil $SU(N_f/2)_1$&
&\hfil $SU(N_f/2)_2$&
&\hfil $SU(N_f/2)_3$&
&\hfil $SU(N_f/2)_4$&
&\hfil $U(1)_e$&
\cr
height3pt
&\omit&
&\omit&
&\omit&
&\omit&
&\omit&
&\omit&
\cr
\noalign{\hrule height 1.1pt}
height3pt
&\omit&
&\omit&
&\omit&
&\omit&
&\omit&
&\omit&
\cr
&\hfil $Q^1$&
&\hfil $N_f/2$&
&\hfil $1$&
&\hfil $1$&
&\hfil $1$&
&\hfil $+1$& 
\cr
height3pt
&\omit&
&\omit&
&\omit&
&\omit&
&\omit&
&\omit&
\cr
\noalign{\hrule}
height3pt
&\omit&
&\omit&
&\omit&
&\omit&
&\omit&
&\omit&
\cr
&\hfil $\Lambda^2$&
&\hfil $1$&
&\hfil $N_f/2$&
&\hfil $1$&
&\hfil $1$&
&\hfil $-1$& 
\cr
\noalign{\hrule}
height3pt
&\omit&
&\omit&
&\omit&
&\omit&
&\omit&
&\omit&
\cr
&\hfil $\tilde\Lambda_1$&
&\hfil $1$&
&\hfil $1$&
&\hfil $\bar{N_f/2}$&
&\hfil $1$&
&\hfil $-1$&
\cr
\noalign{\hrule}
height3pt
&\omit&
&\omit&
&\omit&
&\omit&
&\omit&
&\omit&
\cr
&\hfil $\tilde Q_2$&
&\hfil $1$&
&\hfil $1$&
&\hfil $1$&
&\hfil $\bar{N_f/2}$&
&\hfil $+1$&
\cr
\noalign{\hrule}
height3pt
&\omit&
&\omit&
&\omit&
&\omit&
&\omit&
&\omit&
\cr
&\hfil $X$&
&\hfil $1$&
&\hfil $1$&
&\hfil $1$&
&\hfil $1$&
&\hfil 0&
\cr
\noalign{\hrule}
height3pt
&\omit&
&\omit&
&\omit&
&\omit&
&\omit&
&\omit&
\cr
&\hfil $\Lambda_X$&
&\hfil $1$&
&\hfil $1$&
&\hfil $1$&
&\hfil $1$&
&\hfil 0&
\cr
&\omit&
&\omit&
&\omit&
&\omit&
&\omit&
&\omit&
\cr
}\hrule height 1.1pt
}
$$
}

\centerline{\sl Table 1: The quantum numbers of the light states of the $A_k$ electric model.} 

\bigskip

To implement the construction of \KutasovFFL, we compactify the theory on a two-torus, and turn on a magnetic field (and an appropriate auxiliary $D$ field to preserve $(0,2)$ SUSY) for $U(1)_e$, the global symmetry that assigns charge $+1$ to half of the $N_f$ flavors of $Q$, $\tilde Q$, and $-1$ to the other half.\foot{Thus, we take $N_f$ to be even, as in \KutasovFFL; a similar construction can be implemented for odd $N_f$.} The adjoint field $X$ is not charged under $U(1)_e$, due to the presence of the superpotential \epot.

The effects of the background $B$ and $D$ on all fields other than $X$ were discussed in \KutasovFFL, and will not be reviewed here. Since $X$ is uncharged under $U(1)_e$, it gives rise at long distances to a $(0,2)$ adjoint chiral superfield that we shall also denote by $X$, and an adjoint Fermi superfield $\Lambda_X$. The resulting spectrum and transformation properties of the low energy fields under the global symmetry are given in Table 1.

The four dimensional superpotential \epot\ gives rise in the low energy theory to a $(0,2)$ superpotential (\ie\ an interaction of the form $\int d\theta^+\WW_{(0,2)}+{\rm c.c.}$)
\eqn\epottwo{
\WW_{(0,2)}\sim \Lambda_X X^k~.
}
In addition to the fields in Table 1, the theory contains the adjoint superfields $\Sigma, \Upsilon$, which arise from the reduction of the four dimensional vector superfield, and are singlets under all the symmetries listed in Table 1.  

The classical Coulomb moduli space parametrized by the eigenvalues of $\Sigma$ is replaced in the quantum theory by a discrete set of vacua labeled by an integer $N$ \KutasovFFL. For given $N$, the low energy theory is a direct product of theories with gauge groups $U(N)$ and $U(N_c-N)$, and matter fields in the representations listed in Table 1. The parameter $N$, which naively takes value in $[0,N_c]$, actually satisfies in general more stringent constraints coming from stability of the vacuum in the quantum theory. These constraints can be determined in the same way as in four dimensions, by deforming the superpotential \epottwo\ to a more general one, $\Lambda_X \WW'(X)$,  with the same large $X$ behavior as \epottwo. For generic $\WW$, the low energy theory splits in the infrared into decoupled theories of the sort studied in \KutasovFFL. Since these theories do not have a vacuum when the number of colors is larger than that of flavors (which, as explained in \KutasovFFL, is $N_f/2$), the situation is very similar to the one in four dimensions, and we conclude that the numbers of colors and flavors must satisfy the constraint $N\le kN_f/2$ for a supersymmetric vacuum to exist. Taking into account a similar constraint from the $U(N_c-N)$ theory, we conclude that the parameter $N$ must lie in the range
\eqn\nrange{
\max (0, N_c - \frac k 2 N_f) \leq N \leq \min (N_c, \frac k 2 N_f)~.
}
This condition implies that the quantum theory has supersymmetric vacua if and only if $kN_f\ge N_c$, just like its four dimensional ancestor \stabv. The number of discrete branches of quantum vacua (different values of $N$ which satisfy \nrange) for given $N_f,N_c$ in this range is
\eqn\nvac{N_{\rm br} = \cases{kN_f-N_c+1&$\;\;{\rm for}\;\;\frac k2N_f\le N_c$\cr
N_c+1&$\;\;{\rm for}\;\;\frac k2 N_f\ge N_c.$\cr}}
For $k=1$, \nrange, \nvac\ reduce to the corresponding results in \KutasovFFL. In that case, the expression for $N_{\rm br}$ is invariant under the transformation $N_c\to N_f-N_c$, which is a manifestation of  Seiberg duality in the two dimensional $(0,2)$ theory. For general $k$, \nvac\ is invariant under $N_c\to kN_f-N_c$, as might be expected from the duality of \refs{\KutasovVE\KutasovNP-\KutasovSS}. 

To calculate the central charges of the two factors in the low energy theory (for given $N$) we again proceed as in \KutasovFFL. Consider \eg\ the theory with gauge group $U(N)$. If the number of flavors is in the range $N_f\ge 2N$, the D-term conditions of the $U(N)$ gauge group allow us to turn on expectation values for the fields $Q^1$, $\tilde Q_2$ that break the whole gauge group. In that case, the central charge receives contributions from two sources. One is the massless components of the fundamentals, which contribute to the central charge the amount $3(N_fN-N^2)$. The second is the adjoint superfield $X$, whose contribution, $3N^2(1-{2\over k+1})$, is familiar from $(2,2)$ supersymmetric Landau-Ginzburg theory (see \eg\ \refs{\MartinecZU,\VafaUU}). Adding the two,  we get 
\eqn\ccharge{
c_R = c_L = 3\left(N_f N - {2N^2\over k+1}\right)\,.}
As mentioned above, this result is valid for $N_f\ge 2N$. In \KutasovFFL\ (\ie\ for $k=1$) this condition was required for stability of the vacuum, and therefore placed no limitation on the discussion. For $k>1$, \nrange\ only implies that $N_f\ge 2N/k$, so there is a range of values of $(N,N_f)$, for which the above derivation of \ccharge\  does not apply. As we shall explain shortly, it seems that the result \ccharge\ is valid in the whole range \nrange, but before providing an argument for that, we want to point out a simple consistency check on it. 

If we increase the number of colors $N$ for fixed $N_f$, eventually the central charge \ccharge\ becomes negative, which is inconsistent with the expected unitarity of the two dimensional theory. However, eq. \nrange\ provides an upper bound on $N$, $N\le kN_f/2$, and it is interesting to check whether the central charge is positive when this bound is satisfied. Substituting $N=kN_f/2$ into \ccharge, we find $c=3kN_f^2/2(k+1)$, which is positive, as should be the case if \ccharge\ is generally valid. 

A way to calculate the central charge of the infrared SCFT which does not rely on our ability to completely Higgs the gauge group is to identify the $U(1)_R$ symmetry that becomes part of the $\NN=2$ superconformal multiplet in the infrared, and use the fact that the $U(1)_R^2$ anomaly is  equal to $c_R/3$. In our case, the $U(1)_R$ charges of $\Sigma,\Upsilon$ should be taken to be equal to one for reasons explained in \KutasovFFL. It is natural to assign charge zero to the fields $Q$, $\Lambda$ in the first four rows of Table 1, since at least some of the $Q$'s can be thought of as parametrizing the quantum moduli space, and at infinity in this space behave as free fields. The R-charges of $X, \Lambda_X$ can be obtained from $c$-extermization \refs{\BeniniCZ, \BeniniCDA}, or from the standard discussion of $(2,2)$ Landau-Ginzburg theories \refs{\MartinecZU,\VafaUU}, and are given by $1/(k+1)$. With these charge assignments, the $U(1)_R^2$ anomaly is given by 
\eqn\ranom{
N_f N - \frac 2{k+1}N^2~,
}
in agreement with \ccharge. 

\vbox{
$$\vbox{\offinterlineskip
\hrule height 1.1pt
\halign{&\vrule width 1.1pt#
&\strut\quad#\hfil\quad&
\vrule width 1.1pt#
&\strut\quad#\hfil\quad&
\vrule width 1.1pt#
&\strut\quad#\hfil\quad&
\vrule width 1.1pt#
&\strut\quad#\hfil\quad&
\vrule width 1.1pt#
&\strut\quad#\hfil\quad&
\vrule width 1.1pt#
&\strut\quad#\hfil\quad&
\vrule width 1.1pt#
&\strut\quad#\hfil\quad&
\vrule width 1.1pt#\cr
height3pt
&\omit&
&\omit&
&\omit&
&\omit&
&\omit&
&\omit&
\cr
&\hfil field&
&\hfil $SU(N_f/2)_1$&
&\hfil $SU(N_f/2)_2$&
&\hfil $SU(N_f/2)_3$&
&\hfil $SU(N_f/2)_4$&
&\hfil $U(1)_e$&
\cr
height3pt
&\omit&
&\omit&
&\omit&
&\omit&
&\omit&
&\omit&
\cr
\noalign{\hrule height 1.1pt}
height3pt
&\omit&
&\omit&
&\omit&
&\omit&
&\omit&
&\omit&
\cr
&\hfil $\lambda_1$&
&\hfil $\bar{N_f/2}$&
&\hfil $1$&
&\hfil $1$&
&\hfil $1$&
&\hfil $-1$& 
\cr
\noalign{\hrule}
height3pt
&\omit&
&\omit&
&\omit&
&\omit&
&\omit&
&\omit&
\cr
&\hfil $q_2$&
&\hfil $1$&
&\hfil $\bar{N_f/2}$&
&\hfil $1$&
&\hfil $1$&
&\hfil $+1$& 
\cr
\noalign{\hrule}
height3pt
&\omit&
&\omit&
&\omit&
&\omit&
&\omit&
&\omit&
\cr
&\hfil $\tilde q^1$&
&\hfil $1$&
&\hfil $1$&
&\hfil ${N_f/2}$&
&\hfil $1$&
&\hfil $+1$&
\cr
\noalign{\hrule}
height3pt
&\omit&
&\omit&
&\omit&
&\omit&
&\omit&
&\omit&
\cr
&\hfil $\tilde \lambda^2$&
&\hfil $1$&
&\hfil $1$&
&\hfil $1$&
&\hfil ${N_f/2}$&
&\hfil $-1$&
\cr
\noalign{\hrule}
height3pt
&\omit&
&\omit&
&\omit&
&\omit&
&\omit&
&\omit&
\cr
&\hfil $\hat X$&
&\hfil $1$&
&\hfil $1$&
&\hfil $1$&
&\hfil $1$&
&\hfil 0&
\cr
\noalign{\hrule}
height3pt
&\omit&
&\omit&
&\omit&
&\omit&
&\omit&
&\omit&
\cr
&\hfil $\Lambda_{\hat X}$&
&\hfil $1$&
&\hfil $1$&
&\hfil $1$&
&\hfil $1$&
&\hfil 0&
\cr
\noalign{\hrule}
height3pt
&\omit&
&\omit&
&\omit&
&\omit&
&\omit&
&\omit&
\cr
&\hfil $(M_j)_1^1$&
&\hfil $N_f/2$&
&\hfil $1$&
&\hfil $\bar{N_f/2}$&
&\hfil $1$&
&\hfil 0&
\cr
\noalign{\hrule}
height3pt
&\omit&
&\omit&
&\omit&
&\omit&
&\omit&
&\omit&
\cr
&\hfil $(\Lambda^M_j)_1^1$&
&\hfil $N_f/2$&
&\hfil $1$&
&\hfil $\bar{N_f/2}$&
&\hfil $1$&
&\hfil 0&
\cr
\noalign{\hrule}
height3pt
&\omit&
&\omit&
&\omit&
&\omit&
&\omit&
&\omit&
\cr
&\hfil $(M_j)_2^2$&
&\hfil $1$&
&\hfil $N_f/2$&
&\hfil $1$&
&\hfil $\bar{N_f/2}$&
&\hfil 0&
\cr
\noalign{\hrule}
height3pt
&\omit&
&\omit&
&\omit&
&\omit&
&\omit&
&\omit&
\cr
&\hfil $(\Lambda_j^M)_2^2$&
&\hfil $1$&
&\hfil $N_f/2$&
&\hfil $1$&
&\hfil $\bar{N_f/2}$&
&\hfil 0&
\cr
\noalign{\hrule}
height3pt
&\omit&
&\omit&
&\omit&
&\omit&
&\omit&
&\omit&
\cr
&\hfil $(M_j)_2^1$&
&\hfil $N_f/2$&
&\hfil $1$&
&\hfil $1$&
&\hfil $\bar{N_f/2}$&
&\hfil +2&
\cr
\noalign{\hrule}
height3pt
&\omit&
&\omit&
&\omit&
&\omit&
&\omit&
&\omit&
\cr
&\hfil $(\Lambda^M_j)^2_1$&
&\hfil $1$&
&\hfil $N_f/2$&
&\hfil $\bar{N_f/2}$&
&\hfil $1$&
&\hfil $-2$&
\cr
height3pt
&\omit&
&\omit&
&\omit&
&\omit&
&\omit&
&\omit&
\cr
}\hrule height 1.1pt
}
$$
}
\centerline{\sl Table 2: The quantum numbers of the light states of the $A_k$ magnetic model.} 

\bigskip

As mentioned above, the four dimensional $A_k$ gauge theory is believed to exhibit a generalization of Seiberg duality \refs{\KutasovVE\KutasovNP-\KutasovSS}. The dual theory is a $U(k N_f - N_c)$ gauge theory with $N_f$ chiral superfields in the (anti) fundamental representation $q_i$, $\tilde q^i$, an adjoint field $\hat X$, and $k$ gauge singlets $M_j$, $j= 1, \cdots, k$, which transform in the bifundamental representation of the $SU(N_f)\times SU(N_f)$ flavor group.  The magnetic superpotential takes the form 
\eqn\mpot{
\WW \sim \sum_{j=1}^k M_j\tilde q \hat X^{k-j}q + \Tr \hat X^{k+1}
}
where we omitted the coefficients of the different terms, since they will not play a role in our discussion. The duality relates electric and magnetic chiral operators, 
\eqn\mmap{
 Q X^{j-1}\tilde Q \leftrightarrow M_j ,\;\;  {\rm Tr} X^j\leftrightarrow {\rm Tr}\hat X^j.
}
Compactification to two dimensions with the background fields for  the symmetry $U(1)_e$ gives rise to a $(0,2)$ supersymmetric theory.  
The spectrum of this theory and transformation properties of its low energy fields  (other than $\Upsilon, \Sigma$) under the global symmetries are given in Table 2. 

The magnetic superpotential \mpot\ reduces in the low energy theory to the $(0,2)$ superpotential 
\eqn\mpottwo{
\WW_{(0,2)} \sim \sum_{j=1}^k \left((M_j)^1_1\tilde q^1 \hat{X}^{k-j} \lambda_1 + (M_j)^2_2 \tilde\lambda^2  \hat{X}^{k-j} q_2  + (\Lambda^j_M)^2_1  \tilde q^1 \hat{X}^{k-j}    q_2 \right) + \Lambda_{\hat X}{\hat X}^k.
}

\noindent
One can study the magnetic theory in two steps. First, turn off the couplings between the magnetic mesons and the rest of the theory. Then,  the magnetic mesons act as free spectators; the rest of the theory becomes identical to the electric theory, with the replacement $N_c\to kN_f-N_c$.  Thus, it splits into discrete vacua labeled by the parameter $\hat N$ that takes value in the range (compare to \nrange)
\eqn\mnvac{
\max (0, \frac k 2 N_f - N_c) \leq \hat N \leq \min(kN_f - N_c, \frac k2 N_f).
}
For given $\hat N$ the low energy theory is a direct product of $U(\hat N)$ and $U(kN_f-N_c-\hat N)$ theories. 

We can now couple the magnetic mesons to the resulting theory via the superpotential \mpottwo.  This is not expected to change the vacuum structure \mnvac. As in \KutasovFFL, the relation between $N$ in the electric theory and $\hat N$ in the magnetic one can be obtained by comparing the various $SU(N_f/2)^2$ anomalies. One finds that 
\eqn\gmap{
\hat N = \frac k 2 N_f - N.
}
This is compatible with \nrange, \mnvac, and is related to the fact that the number of branches of the quantum moduli space \nvac\ is duality invariant. 

To calculate the central charge of the low energy theory we need to identify the $U(1)_R$ symmetry that becomes part of the superconformal multiplet in the infrared. To do that, we can proceed as in \KutasovFFL. $\Upsilon, \Sigma$ have R-charge one, and we shall again assume that the diagonal mesons mass up in pairs (\eg\ $(M_j)^1_1$ with $(\Lambda_j^M)^1_1)$ and do not contribute to the $U(1)_R^2$ anomaly. The R-charges of the $(M_j)_2^1$ follow from the part of the operator map \mmap\ which descends to two dimensions,\foot{The $j$ dependence of these R-charges can be understood directly in the magnetic theory. It is due to the fact that the low energy effective Lagrangian has terms that involve the chiral operator $\sum_j M_j \hat X^{k-j}$.}
\eqn\rmj{
R\left[(M_j)_2^1)\right] = \frac{j-1}{k+1}\;.
}
$\hat X, \Lambda_{\hat X}$ again have charge $1/(k+1)$. Finally, we assign to the magnetic quarks $q_2, \tilde q^1$ charge $R_q$, and to $\lambda_1, \tilde\lambda^2$ charge $R_\lambda$. The magnetic superpotential \mpottwo\ implies that $R_q = -R_\lambda$ and 
\eqn\rlambda{
R\left[(\Lambda_j^M)_1^2\right] = 1 - \frac{k-j}{k+1} - 2R_q.}
All R-charges are now determined by $R_q$. To determine it we assume, as in \KutasovFFL, that the $U(1)_R^2$ anomaly in the magnetic theory 
\eqn\gmranom{
N_f\hat N (R_q-1)^2 - N_f \hat N R_q^2 + \sum_{j=1}^{k}\left( \frac{N_f^2}{4}\left[\frac {j-1}{k+1}-1\right]^2 - \frac{N_f^2}{4}\left[1-\frac {k-j}{k+1}-2R_q\right]^2 \right) - \frac 2{k+1} \hat N^2,
}
should coincide with that in the electric theory, \ranom. This gives
\eqn\grq{
R_q = \frac{4N-kN_f}{2kN_f} = \frac{N - \hat N}{k N_f},
}
generalizing the results of \KutasovFFL\ to arbitrary $k$. As a check, the electric theory has a non-vanishing $U(1)_e U(1)_R$ anomaly
\eqn\eranom{
U(1)_e U(1)_R :\; - N N_f.
}
It is easy to verify that the $U(1)_eU(1)_R$ anomaly for the magnetic theory with the above charge assignments agrees with the electric result.

To summarize, we find that the results of \KutasovFFL\ for SQCD generalize naturally to the $A$ series of four dimensional SCFT's found in  \refs{\KutasovVE\KutasovNP-\KutasovSS}. In particular, it appears that the four dimensional duality described in these papers gives rise to a non-trivial identification between different $(0,2)$ models in two dimensions. Some detailed elements of the picture are different for general $k$ than in the $k=1$ case discussed in \KutasovFFL. In particular, for small $N_f$ the electric theory no longer has the property that for large expectation values of the fundamentals the gauge symmetry is completely broken and the theory becomes free. In these cases, it is strongly coupled in the infrared everywhere on the quantum moduli space, and in some cases a better description is the magnetic one. We next turn to the $D$ series where, as we shall see, some additional new features appear.

\newsec{$D$ series}

The $D_{k+2}$ theory is obtained by adding to $\NN=1$ SQCD two chiral superfields in the adjoint representation of $U(N_c)$, $X,Y$,  with the superpotential \rgfp
\eqn\epotdk{
\WW \sim\Tr \left(X^{k+1}+XY^2\right).
}
This theory was studied in \refs{\BrodieVX,\IntriligatorMI}. The stability bound (the analog of \stabv\ for this case) is 
\eqn\stabdk{
N_f \geq \frac{N_c}{3k}.
}
For odd $k$ it can be derived in a way similar to that reviewed above for the $A$ series, by deforming the superpotential \epotdk\ to a generic polynomial with the same large field behavior. For even $k$,  \stabdk\ has not been derived classically, but it is believed to be valid in the quantum theory; see \IntriligatorMI\ for a discussion. 

Upon compactification and turning on the background fields we find a theory with matter content similar to that of the $A_k$ theory (Table 1), with an extra chiral and Fermi multiplet pair $Y$, $\Lambda_Y$, and superpotential
\eqn\epotd{
\WW_{(0,2)} \sim \Lambda_X \left(X^k +  Y^2\right) + \Lambda_Y\{X,Y\}\,.
}
One can repeat the discussion of the $A$ series for this case. Vacua are again labeled by an integer $N$, 
 \eqn\drange{
 \max (0, N_c - \frac{3k}{2}N_f) \leq N \leq \min (N_c, \frac{3k}{2}N_f)\;.
 }
 The number of values that $N$ takes is 
\eqn\nvacdd{N_{\rm br} = \cases{3kN_f-N_c+1&$\;\;{\rm for}\;\;\frac {3k}2N_f\le N_c$\cr
N_c+1&$\;\;{\rm for}\;\;\frac {3k}2 N_f\ge N_c~.$\cr}}
For sufficiently large $N_f$ the gauge group is again generically broken by the expectation values of $Q$, $\tilde Q$, and the central charge can be computed at weak coupling to be 
\eqn\cchargedd{
c_R = c_L = 3\left(N_f N - {N^2\over k+1}\right)\,.}
This result can be alternatively obtained by assigning R-charges 0 to $Q, \Lambda$, 1 to $\Sigma, \Upsilon$, $\frac 1{k+1}$ to $X, \Lambda_X$ and $\frac k{2(k+1)}$ to $Y, \Lambda_Y$. 

A new element in this case is that as we increase $N$ to the maximal value consistent with \drange, $N=3kN_f/2$,  the central charge \cchargedd\ becomes negative at some point. It does so in a region in which the theory is strongly coupled in the infrared, and therefore it is possible that continuing \cchargedd\ there is unwarranted. However, it is not clear from this perspective what does happen in this regime, and in particular why we are running into this problem for the $D$ series and not for the $A$ series. To answer these questions it is useful to employ a dual description.

The four dimensional theory with superpotential \epotdk\ has a dual description found in \BrodieVX. The gauge group is $U(3kN_f - N_c)$; the matter chiral superfields include two adjoints $\hat X, \hat Y$, $N_f$ magnetic quarks $q, \tilde q$, $3kN_f^2$ singlets $(M_{\ell j})^i_{\tilde i}$, $\ell = 1, \cdots, k$, $j = 1,2,3$, $i, \tilde i=1,\cdots, N_f$, and the superpotential
\eqn\spotl{
\WW \sim \Tr \hat X^{k+1} + \Tr \hat X\hat Y^2 + \sum_{\ell = 1}^k \sum_{j=1}^3 M_{\ell j}\tilde q \hat X^{k-\ell}\hat Y^{3-j} q\;.
}
The magnetic mesons correspond to the electric chiral operators
\eqn\mesons{
M_{\ell j} \leftrightarrow \tilde Q X^{\ell -1}Y^{j-1} Q, \qquad \ell= 1,2,\cdots, k;\;\; j = 1,2,3.
}
Applying the procedure of \KutasovFFL\ to this theory, we find a two dimensional $(0,2)$  theory that is similar to the one in \KutasovFFL, and in the previous section. The magnetic mesons $M_{lj}$, which are $N_f\times N_f$ matrices, split into $(N_f/2)\times (N_f/2)$ components $(M_{lj})_\alpha^\beta$, $\alpha,\beta=1,2$, as in \KutasovFFL. The diagonal components,  $(M_{lj})_\alpha^\alpha$, $\alpha=1,2$, are uncharged under $U(1)_e$ and thus give in two dimensions chiral and Fermi multiplets, while the off-diagonal ones give either chiral or Fermi multiplets, depending on the sign of the charge. The magnetic superpotential \spotl\ gives in two dimensions the $(0,2)$ superpotential 
\eqn\epotdm{\eqalign{
\WW_{(0,2)}  &\sim \sum_{\ell=1}^k\sum_{j=1}^3 \left((M_{\ell j})^1_1  \tilde q^1 \hat X^{k-\ell}\hat Y^{3-j} \lambda_1 + (M_{\ell j})^2_2 \tilde\lambda^2 \hat X^{k-\ell}\hat Y^{3-j} q_2 +  (\Lambda^M_{\ell j})^2_1 \tilde q^1 \hat X^{k-\ell}\hat Y^{3-j} q_2 \right)\cr &+ \Lambda_{\hat X} \left(\hat X^k + \hat Y^2 \right)+ \Lambda_{\hat Y}\{\hat X, \hat Y\} \;.
}}
Vacua of the magnetic theory are again governed by an analog of \drange, with $N_c\to 3kN_f-N_c$ and $N\to\hat N$. 
The analog of \gmap\ relating the electric and magnetic vacua obtained from the matching of $[SU(N_f/2)^2]$ anomalies is 
\eqn\nmapd{
 \hat N = \frac{3k}{2}N_f - N\;.
 }
The central charge \cchargedd\ can be obtained by studying the superconformal $U(1)_R$. The R-charge of the $(0,2)$ chiral superfield $M^1_2$ follows from the operator map \mesons,
\eqn\rmjtwo{
R[(M_{\ell j})^1_2] = \frac{2(\ell -1) + k(j-1)}{2(k+1)}.
}
We again assign R-charge $R_q$ to the magnetic quarks, and express the charges of other fields in terms of it using the constraints described in \KutasovFFL, and in the $A$ series discussion in the previous section. For example, the charge of the Fermi superfields $\Lambda^2_1$ that follows from the superpotential \epotdm\ is
\eqn\rlambdatwo{ 
R\left[(\Lambda_{\ell j}^M)_1^2\right] = 1 - \frac{2(k-\ell) + k(3-j)}{2(k+1)} - 2R_q.
}
Plugging in these charges into the $U(1)_R^2$ anomaly, one finds 
 \eqn\ranomd{\eqalign{
U(1)_R^2 &: N_f \hat N (R_q-1)^2 - N_f \hat N R_q^2 + \frac{N_f^2}{4}\sum_{\ell = 1}^k \sum_{j=1}^3 \left(\left[\frac{2(\ell-1) + k(j-1)}{2(k+1)}-1 \right]^2\right. \cr &- \left.\left[1 - \frac{2(k-\ell) + k(3-j)}{2(k+1)}- 2R_q \right]^2\right) - \frac{1}{k+1}\hat N^2.
 }}
 Equating this to the electric result for $c_R/3$  \cchargedd\  gives 
 \eqn\rqd{
 R_q = \frac{4N - 3k N_f}{6kN_f} = \frac{N - \hat N}{3k N_f}.
 }
As in the other examples, the $U(1)_eU(1)_R$ anomaly of the electric theory \eranom\ and the magnetic theory,
\eqn\ermag{
\frac{N_f^2}{2} \sum_{l=1}^k\sum_{j=1}^3 \left[ \frac{2(\ell-1) + k(j-1)}{2(k+1)}   - \frac{2(k-\ell) + k(3-j)}{2(k+1)}- 2R_q   \right]- \hat N N_f = -NN_f~,
}
agree as well. 

It is interesting to revisit in the magnetic language the problem of negative central charge discussed in the electric language after \cchargedd. When $N$ approaches $3kN_f/2$, the magnetic theory satisfies $\hat N\ll N_f/2$. Thus, the gauge dynamics is expected to be weakly coupled there, and the question we should address is what is the role of the Yukawa couplings in \epotdm. In particular, consider the last term on the first line of that equation, the coupling of the gauge singlet Fermi superfields $(\Lambda^M_{lj})^2_1$ to the magnetic quarks. Before coupling the magnetic mesons, the operator to which $(\Lambda^M_{lj})^2_1$ eventually couples,  $\tilde q^1 \hat X^{k-\ell}\hat Y^{3-j} q_2$, has R-charge 
\eqn\rrqqq{R_{lj}={k-l\over k+1}+{(3-j)k\over2(k+1)}\,.}
In order for the coupling to $(\Lambda^M_{lj})^2_1$ to be relevant, this R-charge must satisfy the inequality $R_{lj}<1$. It is easy to see that this inequality is satisfied by roughly half of the operators: all those with $j=3$, and those with $j=2$ and $l>(k/2)-1$. For the rest, our assumption that the R-charge of $\Lambda^M$ is determined by the last term on the first line of \epotdm\ is incorrect. What does happen to these operators depends on their other interaction, but the magnetic description makes it clear that for small $\hat N$ the analysis needs to be modified. 

\newsec{Other cases}

In the previous sections we discussed the $A_k$ and $D_{k+2}$ theories from the ADE classification of 4d $\NN=1$ SQCD theories \rgfp. Here, we would like to briefly comment on the application of the procedure of \KutasovFFL\ to the other examples in \rgfp. 

As before, for sufficiently large $N_f$ the central charge of the low energy $(0,2)$ theory can be calculated by taking the expectation values of the fundamentals to be large and using the fact that the gauge group is then completely broken. This gives
\eqn\cade{
c_R = c_L = 3(N_f N - N^2) + c_{XY} N^2
}
where $c_{XY}$ is the central charge of the corresponding $N=2$ minimal model \refs{\MartinecZU,\VafaUU}, 
\eqn\rgfptwo{
\matrix{
\hat O & 6 \cr
\hat A & 3 \cr
\hat D & 3 \cr
\hat E & 4 \cr
 E_6 & 5/2 \cr
 E_7 & 8/3  \cr
 E_8 & 14/5\,.
 }}
 For the first four entries in \rgfptwo, $c_{XY}\ge 3$, which implies that the central charge \cade\ is positive for all $(N, N_f)$. This is compatible with the expectation that in four dimensions these theories have a stable supersymmetric vacuum for all $N_f$, since as we saw the lower bound on $N_f$  (\eg\ \stabv,\stabdk) is the same in two and four dimensions. For the $E$ series theories (the last three entries in \rgfptwo), $c_{XY}<3$, so positivity of \cade\ places a constraint on $N_f$. Indeed, in these cases one expects the number of flavors for which the four dimensional theory has a supersymmetric vacuum to be bounded from below \IntriligatorMI. In fact, the classification \rgfp\ makes it natural to expect that $E$ series theories satisfy some type of Seiberg duality.
 
In \KutasovYQA\ we conjectured such a duality for the $E_7$ theory. The electric theory is $\NN=1$ SQCD with gauge group $U(N_c)$, two adjoint chiral superfields $X,Y$, and superpotential 
\eqn\epote{
\WW \sim\Tr \left(Y^3 + Y X^3\right).
}
The four dimensional stability bound is $N_f \geq N_c/30$. Upon compactification and turning on the background fields, we find the same matter content as in the $D$ series, but with the superpotential 
\eqn\epote{
\WW_{(0,2)} \sim \Lambda_X \left(X^2Y +XYX+ YX^2\right) + \Lambda_Y(Y^2 + X^3)\,.
}
We expect the vacuum structure of the two dimensional theory to be similar to the previous examples, with the vacua labeled by an integer $N$ that takes value in the range
 \eqn\erange{
 \max (0, N_c - 15N_f) \leq N \leq \min (N_c,\, 15N_f)\,.
 }
 The dual description  \KutasovYQA\ has gauge group $U(30N_f - N_c)$, two adjoints $\hat X, \hat Y$, $N_f$ magnetic quarks $q, \tilde q$, and 30 magnetic mesons $M_j, j = 1, \dots, 30$ corresponding to electric operators $\tilde Q\Theta_jQ$ where the $\Theta_j$ are specified in \KutasovYQA. There is also a magnetic superpotential.
Applying the procedure described above, we find again a two-dimensional (0,2) theory. The $U(1)_R^2$ anomaly in the (0,2) magnetic theory turns out to be
  \eqn\ranome{\eqalign{
U(1)_R^2 &: N_f \hat N (R_q-1)^2 - N_f \hat N R_q^2 + \frac{N_f^2}{4}\sum_{j = 1}^{30} \left[\left(\frac{r_j}{2}-1\right)^2 - \left(1 - \frac{r_j}{2} - 2 R_q\right)^2 \right] - \frac{1}{9}\hat N^2\,,
 }}
 where $r_j, j \in 1, \dots, 30$ are the R-charges of $\Theta_j$ in four dimensions, that can be read off from \KutasovYQA. Matching to the central charge of the electric theory \cade\ then implies that
\eqn\rqe{
 R_q = \frac{N - \hat N}{30 N_f}.
 }
The form of \rqe\ guarantees that the $U(1)_eU(1)_R$ anomaly of the electric and magnetic theories are both equal to $-N N_f$.
 
 As in the $D$ series, there is a region in \erange\ where the central charge \cade, \rgfptwo\ is negative. As there, the electric theory is strongly coupled in that regime, and the problem is resolved by noting that the magnetic superpotential includes some irrelevant couplings. 
 
\bigskip 
\bigskip 
 
\noindent{\bf Acknowledgements}: This work was supported in part by DOE grant DE-FG02-90ER40560, NSF Grant No. PHYS-1066293 and the hospitality of the Aspen Center for Physics, and by the BSF -- American-Israel Bi-National Science Foundation. The work of JL was supported in part by an NSF Graduate Research Fellowship.

\listrefs

\bye